\documentclass[11pt,notitlepage]{article}

\usepackage[utf8]{inputenc}
\usepackage{amsmath}
\usepackage{amsfonts}
\usepackage{graphics}
\usepackage{graphicx}
\usepackage{amssymb}
\usepackage{epigraph}
\usepackage{latexsym}
\usepackage{ifthen}
\usepackage{setspace}
\usepackage{epsfig}
\usepackage{multirow}
\usepackage{booktabs}
\usepackage{array}
\usepackage{float}
\usepackage{verbatim}
\usepackage{rotating}
\usepackage{caption}
\usepackage{subcaption}
\usepackage{comment}
\usepackage{stata}
\usepackage{authblk}
\usepackage{bm}
\usepackage{natbib}
\usepackage{geometry}
\graphicspath{{Figures/}{/}}
\title{Extended multivariate generalised linear and non-linear mixed effects models}
\author[$\dagger$$\star$]{Michael J. Crowther}

\affil[$\dagger$]{University of Leicester, Biostatistics Research Group, Department of Health Sciences, Centre for Medicine, University Road, Leicester, LE1 7RH, UK.}
\affil[$\star$]{michael.crowther@le.ac.uk}
\date{}

\begin{document}

\maketitle

\begin{abstract}
Multivariate data occurs in a wide range of fields, with ever more flexible model specifications being proposed, often within a multivariate generalised linear mixed effects (MGLME) framework. In this article, we describe an extended framework, encompassing multiple outcomes of any type, each of which could be repeatedly measured (longitudinal), with any number of levels, and with any number of random effects at each level. Many standard distributions are described, as well as non-standard user-defined non-linear models. The extension focuses on a complex linear predictor for each outcome model, allowing sharing and linking between outcome models in an extremely flexible way, either by linking random effects directly, or the expected value of one outcome (or function of it) within the linear predictor of another. Non-linear and time-dependent effects are also seamlessly incorporated to the linear predictor through the use of splines or fractional polynomials. We further propose level-specific random effect distributions and numerical integration techniques to improve usability, relaxing the normally distributed random effects assumption to allow multivariate $t$-distributed random effects. We consider some special cases of the general framework, describing some new models in the fields of clustered survival data, joint longitudinal-survival models, and discuss various potential uses of the implementation. User friendly, and easily extendable, software is provided.
\end{abstract}

\section{Introduction}
\label{sec:intro}

Given the current trend in improved availability in both access to data, and volume of data, there is the ever increasing need for flexible, and appropriate statistical modelling techniques, and implementations. Consider the electronic health record (EHR) setting, in which anonymised patient-level data can be obtained. In such settings, we inevitably have a complex hierarchical structure to the data, such as multiple biomarkers measured repeatedly, nested within patients, patients nested within GP practice area, GP practice area nested within geographical regions, and so on. Further challenges include time-dependent effects, and non-linear covariate effects, both of which are arguably commonplace in many settings. Therefore, the need for modelling frameworks which can accommodate such complex structures is paramount.

An area of biostatistics that has received remarkable attention is that of joint modelling. Predominantly, this has been in the setting of joint longitudinal-survival models \citep{Gould2015a}, but there has also been substantial work in joint modelling of multivariate data, not including survival outcomes \citep{Goldstein2009,Macdonald-Wallis2012,Sayers2017}. In essence, a model is specified for each outcome, with some form of sharing between outcome models, often done through shared or correlated random effects. Commonplace in joint longitudinal-survival models is linking the `current value' of a continuous longitudinal outcome, directly to survival, through its expected value conditional on subject-specific random effects. Alternatives include transformations of the current value, for example its gradient or rate of change, or its integral to give a cumulative measure. Arguably, these are clinically plausible ways to link such outcomes in many settings, and give us interpretable association parameters, irrespective of how complex the longitudinal model specification may be (such as when using splines).

The aim of this article is to develop an extended framework for the analysis of multivariate, multilevel data, in particular, providing more flexibility in how outcomes can be linked, regardless of the distribution chosen for each particular outcome, and therefore extending multivariate generalised linear mixed effects (MGLME) models. Such models are generally estimated by maximising the marginal likelihood, integrating out the unobserved random effects at each level of the model \citep{Rabe-Hesketh2004}. Alternatives include expectation-maximisation approaches, and quasi-likelihood approaches such as penalised quasi-likelihood (PQL) \citep{Tuerlinckx2006,Rasbash2009}. In this article we concentrate on the Newton-Raphson approach to maximising the marginal likelihood.

A crucial challenge in the estimation of MGLME models is integrating out the random effects at each level. A number of integration techniques have been proposed, and extensively studied, including both non-adaptive and fully adaptive Gauss-Hermite quadrature, Monte Carlo integration, and importance sampling \citep{Tuerlinckx2006,Pinheiro1995,RHgllamm}. The default choice of the core software packages for fitting MGLME models, such as {\tt nlme} in R, {\tt PROC NLMIXED} in SAS, and {\tt gsem} in Stata, is adaptive Gauss-Hermite quadrature (AGHQ). A criticism of AGHQ is that it's computational burden grows exponentially as the dimension of the random effects, at a particular level, grows. This generally makes it infeasible for use in the high dimensional random effects setting. An alternative which has been proposed for such a setting is Monte Carlo integration (MCI), which involves making random draws from the random effects distribution (with mean and variance elements at their current estimates). This is appealing as the number of draws doesn't necessarily have to increase with the addition of more random effects. Indeed the appeal of MCI is in its simplicity, and MCI has been employed in a wide variety of areas \citep{McCulloch1997,Lin2002}. All one needs to do is randomly sample from a distribution, and it can be specified as your random effect distribution. This allows a convenient and easily implementable way to assume, for example, multivariate $t$-distributed random effects. This can be considered a sensitivity analysis for all models used within this framework. We consider this extension in this paper, in particular using quasi-Monte Carlo methods to reduce the number of samples required \citep{Drukker2006}. 

Furthermore, consider a three level model, where often we may only have one random effect at the highest level, many at level 2, and therefore may wish to consider level-specific random effect distributions, which allow a sensitivity analysis to be conducted at each level. This then motivates level-specific integration techniques, such as using the gold standard AGHQ at level 1 if we have a normally distributed random effect, and MCI at level 2 if we have multivariate $t$-distributed random effects, to provide a more usable implementation. We also consider these extensions in this paper.

The paper is organised as follows. In Section \ref{sec:methods1}, we describe the general likelihood, and describe numerical integration techniques for evaluating the likelihood, proposing level-specific random effect distributions and integration techniques. We then introduce our extended complex linear predictor, and an extensive range of possible distributional models for an outcome. The range includes the standard distributions, but also more general distributions, such as non-linear specifications, and general survival models, allowing for delayed entry and censoring. In Section \ref{sec:cases} we describe some special cases of the framework, each of which can be considered a new development in the literature, including general level flexible parametric survival models, relative survival models, joint frailty models, joint longitudinal-survival models, non-linear mixed effect models, and general hazard models. Throughout the article, we will illustrate the accompanying software package, through example syntax, to illustrate how easy it is to apply the models described in this article, and indeed extend them. Finally, in Section \ref{sec:disc} we conclude with a discussion, including some directions for future research.

\section{Multivariate generalised linear and non-linear mixed effects models}
\label{sec:methods1}

\subsection{Likelihood}
We begin with a single level model with $i=1,\dots,n$ response variables, and therefore can write the conditional joint density function for a given observation as
\begin{align}
	p(\bm{y} | \bm{x}, \bm{b}, \bm{\beta}) = \prod_{i=1}^{n}  p_{i}(y_{i}| \bm{x}, \bm{b}, \bm{\beta}) \nonumber
\end{align}
where $\bm{x}$ is a vector of explanatory variables, $\bm{b}$ is a vector of random effects, and $\bm{\beta}$ is our overall parameter vector. Extending this to a two-level model follows, we have the cluster-specific contribution to the likelihood,
\begin{align}
	p(\bm{y} | \bm{x}, \bm{b}, \bm{\beta}) =  \prod_{i=1}^{n} \prod_{j=1}^{t_{i}}  p_{i}(y_{ij}| \bm{x}, \bm{b}, \bm{\beta})  \label{eqn:like2}
\end{align}
where $t_{i}$ is the number of observations within the $i$th cluster. The log likelihood is obtained by integrating out the unobserved random effects, to obtain a marginal log likelihood,
\begin{align}
  ll(\bm{\beta}) = \log \int_{\mathcal{R}^{r}} p(\bm{y} | \bm{x}, \bm{b}, \bm{\beta}) \phi(\bm{b} | \Sigma_{\bm{b}} ) \text{ d} \bm{b}
  \label{eqn:ll1}
\end{align}
where $\mathcal{R}^{r}$ is the $r$-dimensional space, with each dimension spanning the real number line, and $r$ the dimension of the random effects $\bm{b}$. We assume $\phi()$ is multivariate normal density for $\bm{b}$, with mean vector $\bm{0}$ and variance-covariance matrix $\Sigma_{\bm{b}}$. Equation (\ref{eqn:like2}) can be expanded with further levels of nesting, with $\Sigma_{\bm{b}}$ becoming block diagonal, with a block for each level. Alternatively, following \cite{Rabe-Hesketh2005}, exploiting conditional independence among level ($l-1$) units given the random effects $\bm{B}^{l}$, where $\bm{B}^{l}=(\bm{b}^{l},\dots,\bm{b}^{L})$, we can write a general level model, with the log likelihood for a unit at the highest level $L$, as,
\begin{align}
	ll_{Li} &= \log \int \phi(\bm{b}^{(L)} | \bm{\Sigma}^{(L)}) \prod p^{(L-1)}(\bm{y} | \bm{x}, \bm{b}^{L}, \bm{\beta}) \text{ d}\bm{b}^{(L)} \nonumber \\
	&=\log \int \phi(\bm{b}^{(L)} | \bm{\Sigma}^{(L)}) \exp \left[\sum \log p^{(L-1)}(\bm{y} | \bm{x}, \bm{b}^{L}, \bm{\beta}) \right]\text{ d}\bm{b}^{(L)} \label{eqn:ll21}
\end{align}
where, for $l=2,\dots,L$
\begin{align}
	p^{(l)}(\bm{y} | \bm{x}, \bm{B}^{l+1}, \bm{\beta}) &= \int \phi(\bm{b}^{(l)} | \bm{\Sigma}^{(l)}) \prod p^{(l-1)}(\bm{y} | \bm{x}, \bm{B}^{l}, \bm{\beta}) \text{ d}\bm{b}^{(l)} \nonumber \\
	&=\int \phi(\bm{b}^{(l)} | \bm{\Sigma}^{(l)}) \exp \left( \sum \log p^{(l-1)}(\bm{y} | \bm{x}, \bm{B}^{l}, \bm{\beta}) \right) \text{ d}\bm{b}^{(l)}
	\label{eqn:ll22}
\end{align}
We will refer to this second log likelihood formulation as a nested log likelihood in the remainder of the article. In order to calculate our marginal log likelihood, we must numerically integrate the random effects at each level. 

\subsection{Estimation}
\label{sec:NI}
Throughout this article, we use direct maximisation of the marginal log-likelihood utilising the Newton-Raphson algorithm, with score and Hessian elements calculated using finite differences \citep{Gouldml}. The challenge in evaluating a general likelihood such as that in Equation (\ref{eqn:ll1}) lies in integrating out the random effects at each level. Widely considered the numerical integration technique of choice, we apply mean-variance fully adaptive Gauss-Hermite quadrature as our default integration technique. Applying AGHQ to Equations (\ref{eqn:ll21}) and (\ref{eqn:ll22}), we have,
\begin{align}
	ll_{Li} &\approx \log \left[ \sum_{q_{r_{L}}=1}^{Q} w_{q_{r_{L}}} \dots \sum_{q_{1}=1}^{Q} w_{q_{1}} \exp \left(\sum \log p^{(L-1)}(\bm{y} | \bm{x}, a_{q_{1}},\dots,a_{q_{r_{L}}}, \bm{B}^{(L)}) \right) \right]
\end{align}
and
\begin{align}
	p^{(l)}(\bm{\beta} | \bm{B}^{(l+1)}) &\approx \sum_{q_{r_{l}}=1}^{Q} w_{q_{r_{l}}} \dots \sum_{q_{1}=1}^{Q} w_{q_{1}} \exp \left( \sum \log p^{(l-1)}(\bm{y} | \bm{x}, a_{q_{1}},\dots,a_{q_{r_{l}}}, \bm{B}^{(l)}) \right)
\end{align}
where $\bm{a}_{\bm{q}}$ and $\bm{w}_{\bm{q}}$ are appropriate nodes and weights, respectively, with $Q$ the number of quadrature points per dimension. The adaptive nature of the approximation greatly improves accuracy compared to its non-adaptive version, meaning a greatly reduced number of quadrature nodes can be used per dimension. We advise using an increasing number of AGHQ points to ensure an accurate approximation is obtained.

A criticism of AGHQ is that it doesn't scale well in the presence of many random effects, predominantly with respect to its computational burden growing exponentially with every additional random effect. For example, if we conduct 7-point quadrature, with 2 random effects, we evaluate our likelihood at $7^2=49$ locations. If we had 6 random effects, this would result in $7^6=117,649$ evaluations. This has motivated alternative techniques.

\subsubsection{Monte Carlo integration}
An alternative to AGHQ is that of Monte Carlo (MC) integration \citep{Tuerlinckx2006}. Consider a likelihood function of a two level model, where we need to integrate out the random effects.
\begin{align}
	ll(\bm{\beta}) = \log \int_{\mathcal{R}^{r}} p(\bm{y} | \bm{x}, \bm{b}, \bm{\beta}) \phi(\bm{b} | \Sigma_{\bm{b}} ) \text{ d} \bm{b} \nonumber
\end{align}
A finite sample approximation to this integral is obtained by sampling $M$ random draws from $N(0,\bm{\Sigma}_{b})$, and computing the sample mean
\begin{align}
	ll(\bm{\beta}) \approx \frac{1}{M} \sum_{m=1}^{M} p(\bm{y} | \bm{x}, \bm{b}_{m}, \bm{\beta}) \nonumber
\end{align}
The important thing to note is $M$ doesn't have to increase when extra random effects are added. This makes MC integration attractive in the presence of high dimensional random effects, in comparison to quadrature. However, $M$ must be sufficiently large to ensure an accurate approximation. Variance reduction techniques can be employed to reduce the required $M$, such as antithetic sampling \citep{Hammersley1956}, but in this article we make use of Halton sequences \citep{Drukker2006}, which are based on primes, and have been shown to greatly reduce the number of draws required. Similarly to Gauss-Hermite quadrature, MC integration can be improved by an adaptive procedure just like AGHQ, resulting in an importance sampling approximation \cite{Tuerlinckx2006}.

Since MC integration simply relies on being able to sample from our proposed random effects distribution, $\phi(\bm{b})$, there is no restriction on the distribution that we choose (though we assume it has mean zero on the scale of the linear predictor). Therefore, we can relax the normality assumption placed on the random effects, by using for example a multivariate $t$-distribution, and simply sampling from that instead \citep{Hofert2013}. Once more this opens up a broader range of models, or indeed a sensitivity analysis to a model assuming normally distributed random effects \citep{Lange1989}.

\subsubsection{Level-specific integration techniques and random effect distributions}

The methods described above all assume the same distributional family for the random effects, across all levels of a model. In this section we discuss how to relax this assumption. Consider,
\begin{align}
	ll_{Li} &= \log \int \phi_{L}(\bm{b}^{(L)} | \bm{\Sigma}^{(L)}) \prod p^{(L-1)}(\bm{y} | \bm{x}, \bm{b}^{L}, \bm{\beta}) \text{ d}\bm{b}^{(L)} \nonumber \\
	&=\log \int \phi_{L}(\bm{b}^{(L)} | \bm{\Sigma}^{(L)}) \exp \left[\sum \log p^{(L-1)}(\bm{y} | \bm{x}, \bm{b}^{L}, \bm{\beta}) \right]\text{ d}\bm{b}^{(L)} \label{eqn:ll31}
\end{align}
where, for $l=2,\dots,L$
\begin{align}
	p^{(l)}(\bm{y} | \bm{x}, \bm{B}^{l+1}, \bm{\beta}) &= \int \phi_{l}(\bm{b}^{(l)} | \bm{\Sigma}^{(l)} \prod p^{(l-1)}(\bm{y} | \bm{x}, \bm{B}^{l}, \bm{\beta}) \text{ d}\bm{b}^{(l)} \nonumber \\
	&=\int \phi_{l }(\bm{b}^{(l)} | \bm{\Sigma}^{(l)}) \exp \left( \sum \log p^{(l-1)}(\bm{y} | \bm{x}, \bm{B}^{l}, \bm{\beta}) \right) \text{ d}\bm{b}^{(l)}
	\label{eqn:ll32}
\end{align}
where $\phi_{l}(\bm{b}^{(l)} | \bm{\Sigma}^{(l)})$ for $l=2,\dots,L$, now represents the distribution of the random effects at the $l$th level. This formulation now allows us to specify different distributions at each level. For example, let's consider a three-level model, with a random intercept at the highest level, and 4 random effects at level 2. Our standard approach would be to assume normally distributed random effects at levels 2 and 3, but we can now investigate the robustness to this assumption, not only by using $t$-distributed random effects for all levels, but investigating level specific sensitivity to the assumption, by first assuming $t$-distributed random effect at level 3 and normal distributed random effects at level 2, and vice versa. This provides both an effective method of sensitivity analysis, but also a more robust multilevel analysis.

Assuming different random effects distributions at different levels raises the issue of which integration techniques can be applied effectively. It is arguably only sensible to use Monte Carlo integration when $L$=2, as the computational burden increases dramatically. Therefore, when using level specific distributions, we propose using level specific integration techniques. Consider the example of a three-level model, where often we would only have a random intercept at the highest level, with many random effects at level 2. In this situation, we suggest using AGHQ at the highest level, but employ MCI at level 2 so as to accommodate the possibly many random effects at that level.

\subsection{Linear predictors}
The standard linear predictor for a general level model can be written as follows,
\begin{align}
  \eta = \bm{X} \bm{\beta} + \sum_{l=2}^{L} \bm{X}^{l} \bm{b}^{l}
  \label{eqn:lp}
\end{align}
where subscripts are omitted for ease of exposition. We have $\bm{X}$ our vector of covariates, which could vary at any level, with associated fixed effect coefficient vector $\bm{\beta}$, and $\bm{X}^{l}$ the vector of covariates with random effects $\bm{b}^{l}$ at level $l$.

In an attempt to maintain substantial generality in model specification, we extend the linear predictor shown in Equation (\ref{eqn:lp}) to a highly flexible specification, which forms the basis for the extended framework. It consists of components, indexed by $r$, and each $r$th component consists of elements, indexed by $s$. These of course can vary between parameters. For simplicity, in the exposition below, we assume each response variable has one parameter with a complex linear predictor, with the remaining parameters assumed scalar. We relax this restriction in Section \ref{sec:nl}. Therefore, for the $i$th response variable, we define the linear predictor $\eta_{i}$ as,
\begin{align}
	\eta_{i}	= g_{i}(E[y_{i} | \bm{X}, \bm{b}]) =  \sum_{r=1}^{R_{i}} \prod_{s=1}^{S_{ir}} \psi_{irs} \label{eqn:lp2}
\end{align}
where $g_{i}()$ is the link function for the $i$th response, $\mu_{i} = E[y_{i} | \bm{X}, \bm{b}] = g^{-1}(\eta_{i})$ is the expected value for the $i$th response variable $y_{i}$. We suppress dependence on covariates and random effects in the notation, for ease of exposition. We have $\psi_{irs}$, which is the $s$th functional element, for the $r$th component, for the $i$th response, with $r=1,\dots,R_{i}$ and $s=1,\dots,S_{ir}$. Some particular cases that $\psi_{irs}$ can take are,
\begin{align}
	\psi_{irs} &= X_{irs}(t)
\end{align}
where $X_{irs}$ is a possibly time-dependent covariate, or indeed could be a simple vector of 1's,
\begin{align}
	\psi_{irs} &= \beta
\end{align}
where $\beta$ is a parameter to be estimated,
\begin{align}
	\psi_{irs} &= b
\end{align}
where $b$ is a latent variable, with level and unit indexes omitted for simplicity, commonly assumed to be normally distributed. All random effects at a particular level are stacked and assumed multivariate normal, by default. We further have,
\begin{align}
	\psi_{irs} &= q(t)
\end{align}
where $q(t)$ is some function of time $t$, such as a linear function, or something more complex such as fractional polynomials or splines, to allow modelling of time-dependent effects,
\begin{align}
	\psi_{irs} &= d_{rs}(E[y_{j} | \bm{X}, \bm{b}]), \qquad \qquad\text{where } j=1,\dots,k, j\neq i
\end{align}
where $d_{rs}()$ is a general function, which could be equivalent to the appropriate link function $g_{j}()$ or its inverse $g_{j}^{-1}()$, or something more complex, such as the first or second derivative with respect to $t$, or the integral of the expected value up to time $t$, when a response is modelled as a function of time. Examples of these include the joint longitudinal-survival framework \citep{Gould2015a}. Such a general structure allows the expected value of one outcome to be dependent on the expected value of another, with even further levels of nesting if so desired.

The combination of the elements to form components can form highly flexible model specifications for use in numerous settings. Some examples will be shown in Section \ref{sec:cases}.

\subsection{Distributional choice}
In this section we discuss distributional models for $p_{i}(y_{i}| \bm{x}, \bm{b}, \bm{\beta})$. A variety of standard distributional choices can be used in combination with the complex linear predictor described in the previous section. We consider the Gaussian distribution, where,
\begin{align}
  p_{i}(y_{i}| \bm{x}, \bm{b}, \bm{\beta}) = \frac{1}{\sqrt{2\pi} \sigma} \exp \left(\frac{(y_{i}-\mu_{i})^2}{2\sigma^2}\right) \nonumber
\end{align}
which assumes an identity link function. We have the Poisson distribution,
\begin{align}
  \log p_{i}(y_{i}| \bm{x}, \bm{b}, \bm{\beta}) = -\mu + y_{i} \log \mu_{i} - \log \Gamma (_{i}+1) \nonumber
\end{align}
which assumes a log link function. We have the Bernoulli distribution,
\begin{align}
  \log p_{i}(y_{i}| \bm{x}, \bm{b}, \bm{\beta}) = y_{i} \log(\mu_{i}) + (1-y_{i}) \log(1-\mu_{i}) \nonumber
\end{align}
where in this case $\mu$ is the probability of success. The default link function is the logit. We have the beta distribution, which is a fractional response model, where the outcome $Y$ is assumed to take a value between 0 and 1.
\begin{align}
  \log p_{i}(y_{i}| \bm{x}, \bm{b}, \bm{\beta}) = \log &\Gamma (s) -\log \Gamma (\mu_{i} s) -\log \Gamma (s-\mu s) \nonumber \\
  &+ (\mu_{i} s -1) \log (y_{i}) + (s-\mu_{i} s - 1)\log (1-y_{i}) \nonumber
\end{align}
where $\mu_{i}$ is the mean of $Y$ and $s$ is a scale parameter fitted on the log scale. The default link function is the logit. We have the binomial distribution, which generalises the Bernoulli distribution as the sum of $k$ independent Bernoulli outcomes. The response $Y$ is assumed to take the values $0,1,\dots,k$,
\begin{align}
  \log p_{i}(y_{i}| \bm{x}, \bm{b}, \bm{\beta}) = \log &\Gamma (k+1) - \log \Gamma (y_{i}+1) -\log \Gamma (k-y_{i}+1) \nonumber \\
  &+ y \log \mu_{i} + (1-y_{i}) \log (1- \mu_{i}) \nonumber
\end{align}
where $\mu$ is the expected value for a single Bernoulli outcome. The default link function is the logit. We have the negative binomial distribution, where
\begin{align}
  \log p_{i}(y_{i}| \bm{x}, \bm{b}, \bm{\beta}) = \log &\Gamma (y_{i} + m) - \log \Gamma (y_{i}+1) - \log \Gamma (y_{i}) \nonumber \\
  &+ m \log p_{i} + y_{i} \log (1-p_{i}) \nonumber
\end{align}
where under mean dispersion we have,
\begin{align}
m &= 1/\alpha \nonumber \\
p_{i} &= \frac{1}{1+\alpha \mu_{i}} \nonumber
\end{align}
where $\mu_{i}$ is the expected value of $Y$ and $\alpha$ is a scale parameter, fitted on the log scale. All the distributions described above are inbuilt to the associated software.

\subsubsection{Non-linear models}

Clearly the above is by no means an exhaustive list; however, we can provide such functionality to allow essentially any function to be specified, to take on that of $p_{i}(y_{i}| \bm{x}, \bm{b}, \bm{\beta})$, and still sync with the general level likelihood, and complex linear predictor from Equation (\ref{eqn:lp2}). In other words, non-linear mixed effects models, such as those used in pharmaco-kinetic or growth modelling studies, are also applicable within this framework \citep{Drikvandi2017}. We show an example of this in Section \ref{sec:nl}.

\subsubsection{Survival models}

Moving to time-to-event distributions, we have the log likelihood for a survival outcome,
\begin{align}
  ll(y_{i}| \bm{x}, \bm{b}, \bm{\beta}) = d_{i} \log p(y_{i}) + (1-d_{i}) \log S(y_{i})
  \label{eqn:lls}
\end{align}
where $p()$ is the density function for $Y$, $S()$ is the survival function, $d_{i}$ is the event indicator (0 censored, 1 event), and $y_{i}$ is the observed survival/censoring time. Here we describe some of the standard parametric survival models used widely in practice. For an exponential distribution, we have
\begin{align}
  S(y) &= \exp (-\lambda y_{i} ) \nonumber \\
  \lambda &= \exp(\eta_{i}) \nonumber
\end{align}
For the Weibull distribution, we have
\begin{align}
  S(y) &= \exp (-\lambda y_{i} ^ \gamma) \nonumber \\
  \lambda &= \exp(\eta_{i}) \nonumber
\end{align}
For the Gompertz distribution, we have
\begin{align}
  S(y) &= \exp (-\lambda \gamma^{-1}(e^{\gamma y_{i}}-1)) \nonumber \\
  \lambda &= \exp(\eta_{i}) \nonumber
\end{align}
For the log normal distribution, we have
\begin{align}
  S(y) &= 1 - \Phi \left( \frac{\log(y_{i})-\eta_{i} }{\sigma}\right) \nonumber
\end{align}
For the log-logistic distribution, we have
\begin{align}
  S(y) &= (1+(\lambda y_{i})^{1/\gamma})^{-1} \nonumber
\end{align}
In terms of time-to-event data, we can also provide close to an exhaustive list, by allowing the analyst to specify a function of the (log) hazard function, which is all that is required, since the cumulative hazard function can be estimated using numerical quadrature \citep{Crowther2014}. The log likelihood in Equation (\ref{eqn:lls}) can be written only in terms of the hazard function
\begin{align}
  ll(y_{i}| \bm{x}, \bm{b}, \bm{\beta}) = d_{i} \log h(y_{i}) - \int_{0}^{y_{i}} h(u) \text{ d}u \nonumber
\end{align}
We can then apply numerical integration, to obtain
\begin{align}
  ll(y_{i}| \bm{x}, \bm{b}, \bm{\beta}) = d_{i} \log h(y_{i}) - \frac{y_{i}}{2}\sum_{q=1}^{Q} w_{q} h(\frac{y_{i}}{2}u_{q}+\frac{y_{i}}{2})
  \label{eqn:llss}
\end{align}
where $\bm{u}$ and $\bm{w}$ are sets of appropriate nodes and weights. Equation (\ref{eqn:llss}) implies that given any function for the (log) hazard, we can estimate a model, syncing with the complex linear predictor. Further details of this general hazard approach can be found in \cite{Crowther2014}. Alternatively, if both the (log) hazard function and cumulative hazard function can be written as closed form functions, then this again will sync with the general likelihood and complex linear predictor described above. For completeness, the cumulative hazard function may be specified, with the hazard function estimated using numerical differentiation. We show an example of this in Section \ref{sec:userhaz}.

\section{Some particular cases}
\label{sec:cases}
In this section we describe some special cases of the general framework, each of which can be considered a new proposal. We illustrate them through example syntax, using the associated Stata package developed with this article, namely, the {\tt megenreg} command.

\subsection{A general level parametric survival model}
\label{sec:genrp}
Crowther et al. (2014) developed the extension of the Royston-Parmar flexible parametric survival model to a two-level mixed effect model \citep{Crowther2014a}. At the highest level, they allowed any number of normally distributed random effects, and illustrated the approach in the analysis of recurrent event data with events nested within patients incorporating a random intercept, and the individual patient data meta-analysis of survival data with patients nested within trials incorporating a random treatment effect. In this article, we can easily generalise to any number of levels, and of course any number of random effects at each level. The Royston-Parmar model uses restricted cubic splines of log time, on the log cumulative hazard scale, with log likelihood defined as,
\begin{align}
  ll(y) = d \left\{ s(\log(y) | \beta_{\bm{k}})+\eta + \frac{\partial s(\log(y) | \beta_{\bm{k}}) + \eta}{\partial y} \right\}- \exp \left[ s(\log(y) | \beta_{\bm{k}})+\eta \right] \nonumber
\end{align}
Further details can be found in the book on Royston-Parmar models by \cite{FPMbook}. We first revisit one of the datasets analysed in Crowther et al. (2014) which consists of 38 patients with kidney disease \citep{McGilchrist1991}. The outcome of interest is infection at the catheter insertion point, with our baseline being time of initial catheter insertion. Patients can experience up to two recurrences of infection, resulting in a total of 58 events. The data set-up is as follows,

\begin{stlog}
	. list patient time infect age female in 1/4, noobs
{\smallskip}
  {\TLC}\HLI{40}{\TRC}
  {\VBAR} patient   time   infect   age   female {\VBAR}
  {\LFTT}\HLI{40}{\RGTT}
  {\VBAR}       1      8        1    28        0 {\VBAR}
  {\VBAR}       1     16        1    28        0 {\VBAR}
  {\VBAR}       2     13        0    48        1 {\VBAR}
  {\VBAR}       2     23        1    48        1 {\VBAR}
  {\BLC}\HLI{40}{\BRC}

\end{stlog}

\noindent where {\tt patient} contains the unique identifier for patients, {\tt time} is the time of infection, {\tt infect} is the event (1) or censoring (0) indicator, {\tt age} contains the patient's age at baseline, and {\tt female} an indicator variable with 0 being male and 1 being female. We can fit the same final model, which used 3 degrees of freedom to model the baseline, adjusting for age and gender, and a normally distributed frailty term, using {\tt megenreg}, as follows,

\begin{stlog}
	. megenreg (time age female M1[patient], family(rp, failure(infect) scale(h) df(3)))
{\smallskip}

\end{stlog}

We first write the command name, followed by a model specification within brackets {\tt ()}. The first variable {\tt time} defines the response variable, followed by elements of the complex linear predictor. Elements, as defined above, are separated by spaces. Random effects are defined by starting with a capital letter, with the cluster indicator variables specified within square brackets {\tt []}. The cluster hierarchy is expressed in a simple way by using $<$ or $>$ (when there are more than 2 levels, examples below). The example shows {\tt patient} at the highest level, and observations nested within. Model specific options follow the comma, defining the distribution with {\tt family()}. As it's a survival distribution, we also specify the {\tt failure()} option, defining which variable contains the event indicator, {\tt infect} in this case. Other model specific options, such as {\tt df()}, are available. In this case, we specify a Royston-Parmar survival model with {\tt rp}, and use three degree of freedom (spline terms), through {\tt df(3)}, and specify {\tt scale(h)} to model on the log cumulative hazard scale. Results are presented in Table \ref{tab:surv1}. We can now conduct a sensitivity analysis, relaxing the normality assumption on the frailty term, by assuming a $t$-distributed frailty. In this case we assume 3 degrees of freedom for the $t$-distribution, as follows,

\begin{stlog}
	. megenreg (time age female M1[patient], family(rp, failure(infect) scale(h) df(3)))
>          , redistribution(t) df(3)
{\smallskip}

\end{stlog}

We use the {\tt redistribution(t)} option to specify $t$-distributed random effects, and {\tt df()} to specify the degrees of the freedom for the $t$-distribution. Results are also presented in Table \ref{tab:surv1}, from which we can see the estimated conditional log hazard ratios for age and gender are fairly robust to the normality assumption of the frailty term.

\begin{table}[!h]
  \centering
  \caption{Results from a Royston-Parmar frailty model with either normal or $t$-distributed random intercept.}
    \begin{tabular}{lrrrrrr}
    \toprule
    \multicolumn{1}{c}{\multirow{3}[3]{*}{Parameter}} & \multicolumn{3}{c}{\multirow{2}[1]{*}{Normal}} & \multicolumn{3}{c}{\multirow{2}[1]{*}{$t$}} \\
          & \multicolumn{3}{c}{}  & \multicolumn{3}{c}{} \\
\cmidrule{2-7}          & \multicolumn{1}{c}{Estimate} & \multicolumn{2}{c}{95\% CI} & \multicolumn{1}{c}{Estimate} & \multicolumn{2}{c}{95\% CI} \\
    \midrule
    age   & 0.007 & 0.018 & 0.032 & 0.007 & 0.015 & 0.029 \\
    female & 1.465 & 2.428 & -0.502 & 1.658 & 2.620 & -0.697 \\
    \_1\_rcs1 & 1.753 & 1.281 & 2.225 & 1.741 & 1.285 & 2.196 \\
    \_1\_rcs2 & 0.355 & 0.025 & 0.684 & 0.373 & 0.040 & 0.706 \\
    \_1\_rcs3 & 0.222 & 0.409 & -0.036 & 0.225 & 0.409 & -0.041 \\
    \_cons & 0.348 & 1.663 & 0.967 & 0.136 & 1.341 & 1.069 \\
    \bottomrule
    \end{tabular}%
  \label{tab:surv1}%
\end{table}%

Now consider a three-level example, required when meta-analysing individual patient recurrent event data, with events nested within patients nested within trials. We illustrate how to fit such a model using {\tt megenreg}, investigating the effect of a binary treatment group, {\tt trt}, as follows,

\begin{stlog}
	. megenreg (time trt M1[trial] M2[trial>patient], family(rp, failure(died) scale(h) df(3)))
{\smallskip}

\end{stlog}

\noindent where we now have a random effect {\tt M1[trial]} at the trial level, and a random effect at the patient level, {\tt M2[trial>patient]}. \cite{Rondeau2006} developed a three-level gamma frailty model, with cubic M-splines for the hazard function. Numerical integration was required to integrate out the random effects at each of the levels. Our implementation goes beyond this, to allow any number of normally distributed random effects at each level, such as a random treatment effect which would often be required in the meta-analysis context at the trial level. This can be fitted as follows,

\begin{stlog}
	. megenreg (time trt M1[trial] trt#M1[trial] M2[trial>patient], family(rp, failure(died) scale(h) df(3)))
{\smallskip}

\end{stlog}

Here, components of the same element are separated by a {\tt \#}. This therefore defines a random intercept and random treatment effect at the trial level, level 3, and a random intercept at the patient level, level 2. By default, an independent variance-covariance structure for the random effects at each level is assumed.

The extension to time-dependent effects can be done through an additional component, for example, we can specify a time-dependent treatment effect by forming an interaction between our treatment variable {\tt trt} and a function of time, say $\log(t)$, as follows,

\begin{stlog}
	. megenreg (stime trt trt#fp(0)@phi M1[id1] M2[id1>id2], family(rp, failure(died) scale(h) df(3)) 
>          timevar(stime))
{\smallskip}

\end{stlog}

We use the convenience function {\tt fp()} to denote a fractional polynomial, in this case of degree 1 with power 0, i.e. log time, and estimate a coefficient (potentially a vector if degree $>$ 1) for it called {\tt phi} using the {\tt @} notation. We then must tell {\tt megenreg} our variable which represents time through {\tt timevar()}. If a function of time is included in the linear predictor of a log time or log hazard scale survival outcome model, then it is automatically detected by the program and the required numerical integration will be applied to calculate the cumulative hazard function, necessary for the likelihood. In the Royston-Parmar model, we are already on the (log) cumulative hazard scale, and therefore numerical differentiation will be applied in this case. An essential capability is to also allow non-linear covariate effects, for example we can add age and age squared very simply.

\begin{stlog}
	. gen age2 = age^2
{\smallskip}
. megenreg (stime trt trt#fp(0)@phi age age2 M1[id1] M2[id1>id2], family(rp, failure(died) scale(h) df(3))
>          timevar(stime))
{\smallskip}

\end{stlog}

Similarly, splines can also be included as standard variables. The examples shown above provide an extremely flexible, and extendable framework in which to model clustered survival data, capturing often seen complexities such as time-dependent effects and non-linear covariate effects.

\subsection{A general level parametric relative survival model}

Relative survival models are used widely, particularly in population based cancer epidemiology \citep{Dickman2004}. Generally they are concerned with modelling the excess mortality in a population with a particular disease, compared to a reference population, which usually comprises national life table information. One of the benefits of the approach is that they do not rely on accurate cause of death information. Concentrating again on the Royston-Parmar model, the log likelihood for a relative survival model can be written as follows,
\begin{align}
  ll(y) = d \log \left\{ \exp \left[s(\log(y) | \beta_{\bm{k}})+\eta \right] \times \frac{\partial s(\log(y) | \beta_{\bm{k}}) + \eta}{\partial y} + h^{*}(y) \right\}- \exp \left[ s(\log(y) | \beta_{\bm{k}})+\eta \right] \nonumber
\end{align}
where $h^{*}(y)$ is the expected mortality in the reference population, at the observations time $y$. Usually this is matched based on age, sex, and calendar year of diagnosis. The two-level relative survival Royston-Parmar model was developed by \cite{Crowther2017}, and can be extended to a general level model here. Such a setting may occur when we have observations nested within hospital districts, nested within geographical areas such as counties. The following fits a three-level relative survival model, assuming the Royston-Parmar model with three degrees of freedom to model the baseline, allowing for a time-dependent treatment effect and non-linear function of age.

\begin{stlog}
	. megenreg (stime trt trt#fp(0)@phi M1[id1] M2[id1>id2], family(rp, failure(died) df(3) scale(h) 
>           bhazard(bhaz)) timevar(stime))
{\smallskip}

\end{stlog}

Where the {\tt bhazard()} option defines the variable which contains the expected mortality at a patient's survival time, in the reference population (often matched on year of diagnosis, sex and age). The above extends relative survival models to any number of levels, random effects, etc..

\subsection{General level joint frailty survival models}
\label{sec:jf}
An area of intense research in recent years is in the field of joint frailty survival models, for the analysis of joint recurrent event and terminal event data. Here we focus on the two most popular approaches, proposed by \cite{Liu2004} and \cite{Mazroui2012}. In both, we have a survival model for the recurrent event process, and a survival model for the terminal event process, linked through shared random effects. In the first specification, we have one random effect which accounts for the correlation between recurrent events, and which is then included in the linear predictor for the terminal event model, multiplied by an association parameter, which estimates the strength of the relationship between the two processes. For example,
\begin{align}
  h_{ij}(y) &= h_{0}(y) \exp (\bm{X}_{1ij} \bm{\beta}_{1} + b_{i}) \nonumber \\
  \lambda_{i}(y) &= \lambda_{0}(y) \exp (\bm{X}_{1i} \bm{\beta}_{2} + \alpha b_{i}) \nonumber \\
\end{align}
where $h_{ij}(y)$ is the hazard function for the $j$th event of the $i$th patient, $\lambda_{i}(y)$ is the hazard function for the terminal event, and $b_{i} \sim N(0,\sigma^2)$. We can fit such a model with {\tt megenreg}, adjusting for treatment in each outcome model,

\begin{stlog}
	. megenreg (rectime trt M1[id1]       , family(rp, failure(recevent) scale(h) df(5))) 
>          (stime   trt M1[id1]@alpha , family(rp, failure(died)     scale(h) df(3)))
{\smallskip}

\end{stlog}

We now have two model specifications, each of which can be completely different. We define a random effect with the same name in each linear predictor, which means only one random effect has been defined, but it is included in both outcome models. By default a random effect has a coefficient of one, but by adding the {\tt @alpha} at the end of the element, means a parameter (called {\tt alpha} in the results table, as it uses the name the user provides) will be estimated instead. This provides a highly convenient way of linking random effects between outcome models.

The second approach consists of two random effects included in the recurrent event model, one of which is also included in the linear predictor of the terminal event model, as follows,

\begin{align}
  h_{ij}(y) &= h_{0}(y) \exp (\bm{X}_{1ij} \bm{\beta}_{1} + b_{1i} + b_{2i}) \nonumber \\
  \lambda_{i}(y) &= \lambda_{0}(y) \exp (\bm{X}_{1i} \bm{\beta}_{2} + b_{2i}) \nonumber \\
\end{align}
where $b_{1i} \sim N(0,\sigma_{1}^2)$ and $b_{2i} \sim N(0,\sigma_{2}^2)$. We give an example of how to fit this model with {\tt megenreg}, this time illustrating how to use different distributions for the recurrent event and terminal event processes,

\begin{stlog}
	. megenreg (rectime trt M1[id1] M2[id1] , family(weibull, failure(recevent))) 
>          (stime   trt M2[id1]         , family(rp, failure(died) scale(h) df(3)))
{\smallskip}

\end{stlog}

This defines two random effects at the {\tt id1} level, which by default have an independent covariance structure. All of the extensions discussed in Section \ref{sec:genrp}, i.e. higher levels, time-dependent effects, non-linear covariate effects and the relative survival extension can be included just as before. The examples above extend the Royston-Parmar model to the joint frailty setting.

\subsection{Generalised multivariate joint longitudinal-survival models}

Joint longitudinal-survival models have been widely developed, but there are many avenues of research where they are lacking in terms of methodological development, and importantly, accessible implementations \citep{Gould2015a}.

\subsubsection{Multiple longitudinal outcomes}

A current topic is the ability to model multiple longitudinal outcomes, and link each to survival \citep{Hickey2016}. For simplicity, we describe a situation of two continuous biomarkers ($i=2,3$), where we wish to investigate the association between each value of the biomarker and the risk of event ($i=1$), commonly known as the current value association structure. We model the survival outcome using a Weibull model. Due to the flexibility of the complex linear predictor, we can further extend this framework, by investigating whether there is an interaction between the two biomarkers, and the risk of event. We make no restriction that the two biomarkers must be measured at the same time, only the commonly made assumption that each time schedule is non-informative. Our model is as follows,
\begin{align}
  Y_{1} &\sim Weib(\lambda,\gamma) \nonumber \\
  Y_{2} &\sim N(\mu_{2},\sigma^{2}_{2}) \nonumber \\
  Y_{3} &\sim N(\mu_{3},\sigma^{2}_{3}) \nonumber
\end{align}
The linear predictor of the survival outcome can be written as follows,
\begin{align}
	\eta_1(t) = \bm{X} \bm{\beta}_{0} + E[y_{2}(t)|\eta_{2}(t)] \beta_{1} + E[y_{3}(t)|\eta_{3}(t)] \beta_{2} + E[y_{2}(t)|\eta_{2}(t)] \times E[y_{3}(t)|\eta_{3}(t)] \beta_{3} \nonumber
\end{align}
where $\bm{X}$ is a vector of baseline covariates, with associated log hazard ratios, $\bm{\beta}_{0}$. We apply this model to a commonly used joint model example dataset in the area of primary biliary cirrhosis. Our survival outcome is time to death from any cause, and we have available two biomarkers, namely, serum bilirubin and prothrombin index, both markers of liver function. We illustrate the data setup required, showing the data for a particular patient,

\begin{stlog}
	. list id logb logp time trt stime died if id==3, noobs
{\smallskip}
  {\TLC}\HLI{65}{\TRC}
  {\VBAR} id       logb       logp      time         trt     stime   died {\VBAR}
  {\LFTT}\HLI{65}{\RGTT}
  {\VBAR}  3   .3364722   2.484907         0   D-penicil   2.77078      1 {\VBAR}
  {\VBAR}  3   .0953102   2.484907   .481875   D-penicil         .      . {\VBAR}
  {\VBAR}  3   .4054651   2.484907   .996605   D-penicil         .      . {\VBAR}
  {\VBAR}  3   .5877866   2.587764   2.03428   D-penicil         .      . {\VBAR}
  {\BLC}\HLI{65}{\BRC}

\end{stlog}

\noindent where {\tt logb} and {\tt logp} are the log of serum bilirubin and log(prothrombin index) $\times 10$, respectively, {\tt time} is the time at which the biomarkers were measured, {\tt trt} is treatment group (either placebo or D-penicillamine), {\tt stime} is the observed event time, and {\tt died} is the event indicator. We can then apply our joint model using,

\begin{stlog}
	. megenreg (stime trt EV[logb]@beta1 EV[logp]@beta2 EV[logb]\#EV[logp]@beta3               
>                                                          , family(weibull, failure(died))) 
>          (logb  fp(1)@l1 fp(1)\#M2[id] M1[id]             , family(gaussian) timevar(time)) 
>          (logp  fp(1)@l2 fp(1)\#M4[id] M3[id]             , family(gaussian) timevar(time)) 
>          , covariance(unstructured)
{\smallskip}

\end{stlog}

\noindent where we assume a random intercept and random linear trend for each of the biomarkers. We specify an unstructured variance-covariance structure through the {\tt covariance()} option. Since we want to link the current value of the biomarkers to survival, we use the {\tt EV[]} syntax to link the expected value of an outcome within the linear predictor of another. They are referred to by using the appropriate response variable name within the square brackets. Since we are linking a time-dependent expected value to survival, we must explicitly use the {\tt fp()} or {\tt rcs()} (for restricted cubic splines) functions when using time, as it must be numerically integrated out in the survival outcome model. For each Gaussian response, we must specify the variable which holds the time of measurements, which does not have to be the same between outcomes, as in this case. Further inbuilt functions include the first and second derivatives of the expected value, with respect to time, using {\tt dEV[]} and {\tt d2EV[]}, respectively, and the integral of the expected value, using {\tt iEV[]}. Results from fitting the above model are presented in Table \ref{tab:mvjm}.

\begin{table}[!h]
  \centering
  \caption{Results of multivariate joint model with either normal of $t$ distributed random effects.}
    \begin{tabular}{rlrrrrrr}
    \multicolumn{2}{c}{\multirow{2}[1]{*}{Parameter}} & \multicolumn{3}{c}{\multirow{2}[1]{*}{Normal}} & \multicolumn{3}{c}{\multirow{2}[1]{*}{$t$}} \\
    \multicolumn{2}{c}{} & \multicolumn{3}{c}{}  & \multicolumn{3}{c}{} \\
    \midrule
    \multicolumn{2}{l}{Survival} & \multicolumn{1}{c}{HR} & \multicolumn{2}{c}{95\% CI} & \multicolumn{1}{c}{HR} & \multicolumn{2}{c}{95\% CI} \\
    \midrule
          & a1    & 2.908 & 2.394 & 3.531 & 2.864 & 2.379 & 3.448 \\
          & a2    & 1.710 & 1.418 & 2.062 & 1.615 & 1.394 & 1.872 \\
          & trt   & 1.094 & 0.762 & 1.569 & 1.012 & 0.711 & 1.439 \\
          & lambda & 0.000 & 0.000 & 0.000 & 0.000 & 0.000 & 0.000 \\
          & gamma & 0.969 & 0.812 & 1.156 & 0.980 & 0.826 & 1.162 \\
          &       &       &       &       &       &       &  \\
    \multicolumn{2}{l}{Longitudinal 1} & \multicolumn{1}{c}{Estimate} & \multicolumn{2}{c}{95\% CI} & \multicolumn{1}{c}{Estimate} & \multicolumn{2}{c}{95\% CI} \\
    \midrule
          & cons1 & 0.491 & 0.377 & 0.605 & 0.280 & 0.211 & 0.349 \\
          & l1    & 0.195 & 0.167 & 0.223 & 0.100 & 0.085 & 0.114 \\
          & sd(resid1) & 0.346 & 0.333 & 0.359 & 0.349 & 0.336 & 0.362 \\
          &       &       &       &       &       &       &  \\
    \multicolumn{2}{l}{Longitudinal 2} & \multicolumn{1}{c}{Estimate} & \multicolumn{2}{c}{95\% CI} & \multicolumn{1}{c}{Estimate} & \multicolumn{2}{c}{95\% CI} \\
    \midrule
          & cons2 & 23.575 & 23.479 & 23.671 & 23.492 & 23.415 & 23.568 \\
          & l2    & 0.226 & 0.192 & 0.260 & 0.159 & 0.139 & 0.179 \\
          & sd(resid2) & 0.746 & 0.718 & 0.775 & 0.759 & 0.732 & 0.786 \\
          &       &       &       &       &       &       &  \\
    \multicolumn{2}{l}{Random effects} & \multicolumn{1}{c}{Estimate} & \multicolumn{2}{c}{95\% CI} & \multicolumn{1}{c}{Estimate} & \multicolumn{2}{c}{95\% CI} \\
    \midrule
          & sd(cons1) & 1.001 & 0.921 & 1.088 & 0.758 & 0.711 & 0.807 \\
          & sd(cons2) & 0.195 & 0.170 & 0.224 & 0.186 & 0.176 & 0.197 \\
          & sd(l1) & 0.721 & 0.640 & 0.811 & 0.469 & 0.418 & 0.526 \\
          & sd(l2) & 0.185 & 0.154 & 0.224 & 0.156 & 0.137 & 0.178 \\
          & corr(cons1,l1) & 0.455 & 0.312 & 0.578 & 0.236 & 0.139 & 0.330 \\
          & corr(cons1,cons2) & 0.596 & 0.487 & 0.687 & 0.720 & 0.629 & 0.791 \\
          & corr(cons1,l2) & 0.508 & 0.346 & 0.641 & 0.111 & -0.056 & 0.271 \\
          & corr(l1,cons2) & 0.437 & 0.251 & 0.592 & 0.105 & -0.010 & 0.216 \\
          & corr(l1,l2) & 0.524 & 0.332 & 0.675 & 0.754 & 0.678 & 0.814 \\
          & corr(cons2,l2) & 0.161 & -0.074 & 0.378 & 0.022 & -0.150 & 0.192 \\
    \end{tabular}%
  \label{tab:mvjm}%
\end{table}%

We can conduct a sensitivity analysis for the above model by using multivariate $t$-distributed random effects, as follows,

\begin{stlog}
	. megenreg (stime trt EV[logb]@beta1 EV[logp]@beta2 EV[logb]\#EV[logp]@beta3 
>                                                          , family(weibull, failure(died)) 
>          (logb  fp(1)@l1 fp(1)\#M2[id] M1[id]             , family(gaussian) timevar(time)) 
>          (logp  fp(1)@l2 fp(1)\#M4[id] M3[id]             , family(gaussian) timevar(time)) 
>          , covariance(unstructured) redistribution(t) df(3)
{\smallskip}

\end{stlog}

\noindent where we specify the {\tt redistribution(t)} and choose a degrees of freedom with {\tt df()}. Results of this model are also shown in Table \ref{tab:mvjm}, indicating some impact on estimates for the longitudinal model for log serum bilirubin (biomarker 1). This provides evidence that the more robust model may be needed.

A further extension, which is simple to incorporate, is to allow time-dependent association parameters, i.e. non-proportional hazards in the association parameter. We can investigate the presence of non-proportional hazards in the association between the second biomarker, and survival, by forming an interaction between the expected value for that outcome, and a function of time, as follows,
\begin{align}
	\eta_1(t) &= \bm{X} \bm{\beta}_{0} + E[y_{2}(t)|\eta_{2}(t)] \beta_{1} + E[y_{3}(t)|\eta_{3}(t)] \beta_{2}(t) \nonumber \\
	&= \bm{X} \bm{\beta}_{0} + E[y_{2}(t)|\eta_{2}(t)] \beta_{1} + E[y_{3}(t)|\eta_{3}(t)] \times \beta_{2} + E[y_{3}(t)|\eta_{3}(t)] \times \beta_{3} \times t \nonumber
\end{align}
This model is implemented using {\tt megenreg} as follows,

\begin{stlog}
	. megenreg (stime trt EV[logb]@beta1 EV[logp]@beta2 fp(0)#EV[logp]@beta3 
>                                                , family(rp, failure(died) df(3)) timevar(stime))
>          (logb  fp(1)@l1 fp(1)\#M2[id] M1[id]   , family(gaussian) timevar(time)) 
>          (logp  fp(1)@l2 fp(1)\#M4[id] M3[id]   , family(gaussian) timevar(time)) 
>          , covariance(unstructured)
{\smallskip}

\end{stlog}

\noindent where {\tt beta2} gives the log hazard ratio for a one unit increase in the biomarker at baseline, $t=0$, and {\tt beta3} is the linear change in {\tt beta2} over log time. The function of time can be as simple or complex as required, for example to allow non-linearity.

\subsubsection{Competing risks}

Extending to a cause-specific competing risks model can also be done in an extremely simple way. Consider the hypothetical situation where we have cause of death information for the PBC example above. We assume patient's either died from PBC or from another cause. We now consider this data structure, where we have {\tt stime} which is a patient's time of death or censoring, but now have cause-specific event indicators, called {\tt diedpbc} and {\tt diedother}

\begin{stlog}
	. list id logb logp time trt stime diedpbc diedother if id==3, noobs
{\smallskip}
  {\TLC}\HLI{76}\HLI{4}{\TRC}
  {\VBAR} id       logb       logp      time         trt     stime   diedpbc   diedother {\VBAR}
  {\LFTT}\HLI{76}\HLI{4}{\RGTT}
  {\VBAR}  3   .3364722   2.484907         0   D-penicil   2.77078         1           0 {\VBAR}
  {\VBAR}  3   .0953102   2.484907   .481875   D-penicil         .         .           . {\VBAR}
  {\VBAR}  3   .4054651   2.484907   .996605   D-penicil         .         .           . {\VBAR}
  {\VBAR}  3   .5877866   2.587764   2.03428   D-penicil         .         .           . {\VBAR}
  {\BLC}\HLI{76}\HLI{4}{\BRC}

\end{stlog}

where {\tt diedpbc} indicates whether a patient died from PBC (1) or was censored at their {\tt stime} (0), and {\tt diedother} indicates whether a patient died from another cause (1) or was censored (0). We can now fit cause-specific models extremely easily as follows,

\begin{stlog}
	. megenreg 
>      (stime trt EV[logb]@a1 EV[logp]@a2        , family(weibull, failure(diedpbc)))
>      (stime trt EV[logb]@a3 EV[logp]@a4        , family(gompertz, failure(diedother)))
>      (logb  fp(1 2)@l1    fp(1)\#M2[id] M1[id]  , family(gaussian) timevar(time))
>      (logp  rcs(df(3))@l2 fp(1)\#M4[id] M3[id]  , family(gaussian) timevar(time)) 
{\smallskip}

\end{stlog}

Parameters {\tt a1} and {\tt a2} gives us the cause-specific association between the current values of serum bilirubin and prothrombin index, respectively, and the hazard of death from PBC. Similarly, {\tt a3} and {\tt a4} give us the association between the current values of the biomarkers and the hazard of death from other causes. The extension to more competing risks follows naturally.

\subsubsection{Recurrent events and a terminal event}

The extensive {\tt frailtypack} in R has recently been extended to fit a joint model of a continuous biomarker, a recurrent event process, and a terminal event \citep{Krol2016,Krol2017}. Here, we illustrate how to fit such a model with {\tt megenreg}, combining the models and syntax shown in Sections \ref{sec:jf} and the above joint longitudinal-survival examples.

\begin{stlog}
	. megenreg (canctime trt EV[logb]@a1 EV[logp]@a2 M5[id]      , family(weibull, failure(canc)))
>          (stime trt EV[logb]@a4 EV[logp]@a5 M5[id]@alpha   , family(gompertz, failure(died))) 
>          (logb  fp(1)@l1 fp(1)\#M2[id] M1[id]               , family(gaussian) timevar(time))
>          (logp  fp(1)@l2 fp(1)\#M4[id] M3[id]               , family(gaussian) timevar(time)) 
{\smallskip}

\end{stlog}

With the addition of a fifth random effect, {\tt M5[id]}, we can account for the correlation between recurrent events, and then link it to the terminal event process, just as above.

\subsubsection{Multistate model}
The final joint model extension comprises of a multistate process, linked to longitudinal outcomes. Once more this is a hypothetical setting, to illustrate the feasibility of fitting such a complex model simply, within the {\tt megenreg} command. We illustrate an illness-death Markov multistate process, whose extension relies on the extension of the survival models to allow for left truncation.

\begin{stlog}
	. megenreg 
>    (canctime    trt EV[logb]@a1 EV[logp]@a2 , family(weibull, failure(canc)))
>    (stimenocanc trt EV[logb]@a4 EV[logp]@a5 , 
>                              family(gompertz, failure(diednocanc) ltrunc(canctime))
>    (stimecanc   trt EV[logb]@a4 EV[logp]@a5 , family(gompertz, failure(diedcanc)))
>    (logb        fp(1)@l1    fp(1)\#M2[id] M1[id]  , family(gaussian) timevar(time))
>    (logp        fp(1)@l2    fp(1)\#M4[id] M3[id]  , family(gaussian) timevar(time)) 
{\smallskip}

\end{stlog}

By specifying a survival model for each transition, we can form the full joint model by linking the longitudinal outcome to survival as before. Each of the transition-specific survival models can be completely different distributions, providing full flexibility.

The above joint models provide a number of methodological extensions. The extensions to more outcomes, outcomes of different types (non-continuous), higher levels, alternative ways to linking outcomes, relative survival, all follow naturally within the framework and implementation.

\subsection{A user-defined distribution}
\label{sec:user}
A central aim of the development of this extended framework is to provide as much generality as possible, whilst still being easily implementable within the associated software. The first of two core aspects of this is the complex linear predictor. We now move to the second, allowing non-standard distributions to be seamlessly incorporated into the estimation engine. As an example, we consider the case of a Gaussian distributed outcome. We must define a simple function which returns the observation level log-likelihood contribution contained in a matrix of real number, which we call {\tt gauss\_logl()}. The function must take one input, which must be of type {\tt transmorphic}, and we call it in this case, {\tt gml}. This is our core object which contains all the information needed about the model to be estimated, and is what is passed to the subroutines. The user must not attempt to edit this object.

\begin{stlog}
	real matrix gauss\_logl(gml)
{\lbr}
     y        = megenreg\_util\_depvar(gml)         //dep. var.
     linpred 	= megenreg\_util\_xzb(gml)            //lin. pred.
     sdre     = exp(megenreg\_util\_xb(gml,1))      //anc. param.
     return(lnnormalden(y,linpred,sdre))         //logl
{\rbr}
{\smallskip}
\end{stlog}

Our function uses the subroutines to extract the response variable, the complex linear predictor, any ancillary parameters (applying an exponential transformation, as the standard deviation is estimated on the log scale), and returns the observation level log likelihood contribution for this outcome, using the official Mata function {\tt lnnormalden()}. We can now use the standard {\tt megenreg} syntax as follows,

\begin{stlog}
	. megenreg (logb time time\#M2[id] M1[id], family(user, loglf(gauss\_logl)) np(1)) 
{\smallskip}

\end{stlog}

We tell {\tt megenreg} that we are using a {\tt user} defined family, and tell it the name of the Mata function using the {\tt loglfunction()} option. We also tell it that we have one ancillary parameter (the standard deviation of the Gaussian distribution) to estimate through the {\tt np()} option. Such an approach means that the user has complete autonomy over the distributional model specified for any particular outcome, whilst still being able to utilise the complex linear predictor provided by the framework. User-defined functions can be used in multiple outcome models, alongside standard distributions available within the command. The example function above is exactly what is implemented in {\tt megenreg} with the {\tt family(gaussian)} option, illustrating how easy it is to define a new distribution through use of the utility functions provided.

\subsection{A non-linear mixed effects example with multiple complex linear predictors}
\label{sec:nl}
In all previous examples above, we have only considered settings where we have a single complex linear predictor, with all other ancillary parameters as scalars. Here, we consider relaxing this restriction, allowing multiple complex linear predictors, within a single outcome model. A central aim of this framework, in the context of the software implementation, is to be able to directly implement non-standard models, which are not available as a freely available package. We consider the example of Murawska et al. (2012). They developed a non-linear joint model estimated with a Bayesian approach (code available from the authors). We show how to directly implement their model as a joint model estimated using maximum likelihood, using {\tt megenreg} with just a few lines of code. They assumed a Gaussian response variable, with multiple non-linear predictors each with fixed effects and a random intercept. The overall non-linear predictor is defined as,

\begin{align}
  f(t) = \beta_{1i} + \beta_{2i} \exp^{-\beta_{3i}t} \nonumber
\end{align}
where each linear predictor was constrained to be positive,
\begin{align}
  \beta_{1i} &=  \exp(X_{1}\beta_{1} + u_{1i}) \nonumber \\
  \beta_{2i} &= \exp(X_{2}\beta_{2} + u_{2i}) \nonumber \\
  \beta_{3i} &= \exp(X_{3}\beta_{3} + u_{3i})   \nonumber
\end{align}
and for the survival outcome
\begin{align}
  \lambda(t) = \lambda_{0}(t)\exp (\alpha_{1}\beta_{1i} + \alpha_{2}\beta_{2i} + \alpha_{3}\beta_{3i})  \nonumber
\end{align}
where either a Weibull or piecewise constant hazard function was chosen for $\lambda_{0}(t)$. In terms of exposition, it's easier to begin with the call to {\tt megenreg} that we use to fit this model. Since we have multiple linear predictors, we consider each one as an outcome model, choosing the first to represent the main linear predictor, plus the survival outcome model which we will assume as Weibull, as follows,

\begin{stlog}
	. megenreg (resp age female M1[id], family(user, llf(nlme\_logl))
>                                            np(1) timevar(time))
>          (age female M2[id],      family(null))                                           
>          (age female M3[id],      family(null))
>          (stime age female EV[1]@alpha1 EV[2]@alpha2 EV[3]@alpha3, 
>                                   family(weibull, failure(died)))
>          , covariance(unstructured)
{\smallskip}

\end{stlog}

In the above syntax, we consider the first model specification as our main one, and hence specify our response variable, the user-defined Mata function to calculate the observation level log likelihood contribution, and the number of ancillary parameters. The second and third model specifications are convenient ways to define additional complex linear predictors. They are flagged as such by defining the {\tt family(null)}, meaning they do not contribute to the log likelihood, but still allow the complex linear predictor to be defined and evaluated. We now define our user-written function to explain further,

\begin{stlog}
	real matrix nlme\_logl(gml, t)
{\lbr}
     y         = megenreg\_util\_depvar(gml)                      //dep.var.
     linpred1 	= exp(megenreg\_util\_xzb(gml))                    //main lin. pred.
     linpred2 	= exp(megenreg\_util\_xzb\_mod(gml,2))              //extra lin. preds
     linpred3 	= exp(megenreg\_util\_xzb\_mod(gml,3))
     sdre      = exp(megenreg\_util\_xb(gml,1))                   //anc. param
     linpred   = linpred1 :+ linpred2:*exp(-linpred3:*t)       //nonlin. pred
     return(lnnormalden(y,linpred,sdre))                       //logl
{\rbr}
{\smallskip}
\end{stlog}

We can use an additional function, {\tt gml\_util\_xzb2()}, to extract a complex linear predictor from a model different to the current one, where current refers to the model in which the {\tt family(user)} was specified. The second option of {\tt gml\_util\_xzb2()} tells it which model linear predictor to extract, in this case we need the second and third. The overall non-linear predictor is then built and passed to the log normal density function, along with the ancillary parameter for the standard deviation of the Gaussian response. This is simply an example of the complete flexibility of the implementation, and of course these few lines of code can be extended to multiple levels, random effects, etc..

\subsection{A user-defined hazard model}
\label{sec:userhaz}

We can extend a general hazard model, such as those developed by \cite{Kooperberg1995} and \cite{Crowther2014} to incorporate the complex linear predictor. Consider a survival model where we wish to model the baseline log hazard function using polynomials of time,
\begin{align}
  h(y) = h_{0}(y) \exp (\eta)
\end{align}
where
\begin{align}
  \log h_{0}(y) = \beta_{0} + \beta_{1} y + \beta_{2} y^2 + \beta_{3} y^3
\end{align}
Such a model can be implemented similarly to that shown in Section \ref{sec:user}. We define a function which takes two inputs, the first being the core object as before, but the second being a {\tt real matrix t}, representing time. We require the second input as time will be integrated out to calculate the cumulative hazard function required in the log likelihood calculation.

\begin{stlog}
	real matrix haz(gml, t)
{\lbr}
     //extract and store the complex linear predictor
     linpred 	= megenreg\_util\_xzb(gml)
     //extract any anciliary parameters
     b        = megenreg\_util\_xb(gml,1),megenreg\_util\_xb(gml,2),megenreg\_util\_xb(gml,3)
     //return the observation level log likelihood
     return(exp(linpred :+ b[1] :* t :+ b[2] :* t:^2 :+ b[3] :* t:^3))
{\rbr}
{\smallskip}

\end{stlog}

\noindent Our function must return the observation level hazard function. We can then fit a recurrent event model, for example, using

\begin{stlog}
	. megenreg (stime trt M1[id], family(user, hfunction(haz)) np(3))
{\smallskip}

\end{stlog}

\noindent Alternatively, if the cumulative hazard has a closed form solution, we can use the {\tt cumhazard()} option as well, to pass a function which calculates the cumulative hazard function. If only the cumulative hazard function is specified, then numerical differentiation is used to calculate the hazard function, to then maximise the log likelihood.

\subsection{Mixed effects for the level 1 variance function}

A recent paper by \cite{Goldstein2017} proposed a two-level model with complex level 1 variation, of the form,
\begin{align}
  y_{ij} &= \bm{X}_{1ij}\bm{\beta}_{1} + \bm{Z}_{1ij}\bm{b}_{1j} + \epsilon_{ij} \nonumber  \\
  \epsilon_{ij} &\sim N(0,\sigma_{e}^{2}) \nonumber  \\
  \log (\sigma_{e}^{2}) &= \bm{X}_{2ij}\bm{\beta}_{2} + \bm{Z}_{2ij}\bm{b}_{2j} \nonumber \\
\begin{pmatrix}
\bm{b}_{1j}\\
\bm{b}_{2j}
\end{pmatrix} & \sim  N\left[\left(\begin{array}{c}
0\\
0
\end{array}\right),\left(\begin{array}{ccc}
\bm{\Sigma}_{b_{1}} &  \\
\bm{\Sigma}_{b_{1}b_{2}} & \bm{\Sigma}_{b_{2}}
\end{array}\right)\right]
\end{align}
They estimated their proposal with a Bayesian approach utilising Markov Chain Monte Carlo methods. Here we can subsume this proposed model within the extended framework proposed in this article, and show how to fit it using {\tt megenreg}, as follows,

\begin{stlog}
	real matrix lev1\_logl(gml)
{\lbr}
     y        = megenreg\_util\_depvar(gml)                //response
     linpred1 = megenreg\_util\_xzb(gml)                   //lin. pred.
     varresid = exp(megenreg\_util\_xzb\_mod(gml,2))        //lev1 lin. pred
     return(lnnormalden(y,linpred,sqrt(varresid)))      //logl
{\rbr}
{\smallskip}

\end{stlog}

\noindent where,

\begin{stlog}
	. megenreg (resp female age age\#M2[id] M1[id], family(user, llf(lev1\_logl)))
           (age female M3[id],                 family(null))                                     
           , covariance(unstructured)
{\smallskip}

\end{stlog}

Now we can extend the level 1 variance function model in a number of ways, simply through the complex linear predictor, for example through extending to higher levels, time-dependent effects and non-linear covariate effects, as discussed in \cite{Goldstein2017}. We can of course also incorporate a non-linear mixed effects model, such as the example in Section \ref{sec:nl}.

\section{Discussion}
\label{sec:disc}

In this article, we have described an extended framework for the analysis of multivariate multilevel data, primarily through a complex linear predictor, which provides substantial flexibility to meet a wide variety of settings. Through the complex linear predictor, we allow seamless development of novel models, and crucially, a way of making them immediately available to researchers through an accessible implementation. We have further showed how the complex linear predictor (or predictors), can be used within a non-linear models, either for a single outcome model, or combined with any number of outcomes and distributions.

In our illustrative examples in Section \ref{sec:cases}, we developed a number of extensions to the Royston-Parmar survival model, namely, an arbitrary number of levels and random effects at each level, the joint frailty setting for the analysis of recurrent event and a terminal event, and the multilevel relative survival extension. Any of the standard parametric survival models can be used instead, or even a user-defined model, such as using splines or fractional polynomials to model the baseline hazard function and time-dependent effects. More research is needed into each particular case to establish performance in realistic settings through simulation studies. We further developed a number of novel contributions to the joint longitudinal-survival literature, including allowing biomarker-biomarker interactions in a multivariate longitudinal joint model, non-proportional hazards in association parameters, and providing an overarching framework for generalised multivariate multilevel longitudinal-survival models, which encompasses anything from a multi-state survival setting, to any number of levels and random effects, and a wide variety of standard and bespoke association structures to link between outcomes.

As discussed earlier in the article, computation time is often a limiting factor with such complex models, especially with high dimensional random effects. On one side, this issue is becoming less of a concern due to the ever increasing power of computers, but conversely, now that we have entered an era of `big data', the size and availability of datasets is also on the increase, which inevitably increases the computational demand. A potential solution is to use efficient sampling weights, such as those utilised in a case-cohort design, meaning we no longer have to analyse the full dataset, with only a minimal loss in precision. This will be investigated in future work, similar in spirit to that of \cite{Rabe-Hesketh2006}. However, often the main challenge with fitting such complex models is the dimension of the random effects. Through the use of Monte Carlo integration, the computational burden can be reduced compared to more commonly used Gauss-Hermite quadrature. This also allowed us to extend all the models described to incorporate $t$-distributed random effects, to provide a more robust model setting, or a sensitivity analysis to the usual normal random effects assumption. Furthermore, we showed how to use level-specific random effect distributions and numerical integration techniques to provide further methods of sensitivity analyses. A further benefit of MCI is that it can be parallelised through the use of stream random number generators, enabling more efficient computations. This will be incorporated into a future version of the software. 

The implementation uses finite differences to calculate the score and Hessian, which requires repeated calls to the log likelihood evaluator. Current work is incorporating analytic derivatives to provide substantial speed gains. Further extensions include allowing for multiple membership and cross clasification, and further development of mixed effect location-scale models \citep{Hedeker2013}.

The methodology described in this article has all been implemented as a freely available software package in Stata. The package is currently being ported to R. Installation instructions and many further examples can be viewd on the accompanying website: mjcrowther.co.uk/software/megenreg.

\section*{Acknowledgments}
I would like to thank Professor Paul Lambert and Adrian Sayers for their very helpful comments on an earlier draft of this paper. MJC is part funded by a MRC New Investigator Research Grant (MR/P015433/1).

\bibliographystyle{biom}
\bibliography{paper_arxiv}

\end{document}